\documentstyle[12pt]{article}

\newcommand{\beq}{\begin{equation}}
\newcommand{\eeq}{\end{equation}}
\newcommand{\beqa}{\begin{eqnarray}}
\newcommand{\eeqa}{\end{eqnarray}}
\newcommand{\sect}[1]{\setcounter{equation}{0}\section{#1}}
\newcommand{\rf}[1]{(\ref{#1})}
\newcommand{\da}{\dot{a}}
\newcommand{\db}{\dot{b}}
\newcommand{\dphi}{\dot{\phi}}
\newcommand{\dda}{\ddot{a}}
\newcommand{\ddb}{\ddot{b}}
\newcommand{\ddphi}{\ddot{\phi}}
\newcommand{\re}{{\rm Re}}
\newcommand{\im}{{\rm Im}}

\begin{document}

\title{The Probability for Primordial Black Holes}
\author{R. Bousso\thanks{\it R.Bousso@damtp.cam.ac.uk} \ and
        S. W. Hawking\thanks{\it S.W.Hawking@damtp.cam.ac.uk}
      \\Department of Applied Mathematics and
      \\Theoretical Physics
      \\University of Cambridge
      \\Silver Street, Cambridge CB3 9EW
       }
\date{DAMTP/R-95/33
    \\gr-qc/9506047}

\maketitle

\begin{abstract}

We consider two quantum cosmological models with a massive
\linebreak scalar field:
an ordinary Friedmann universe and a universe containing primordial
black holes.
For both models
we discuss the complex solutions to the Euclidean Einstein equations.
Using the probability measure obtained from the Hartle-Hawking no-boundary
proposal,
we find that the only unsuppressed black holes start at the Planck size but
can grow with the horizon scale during the roll down of the scalar field
to the minimum.

\end{abstract}

\pagebreak

\sect{Introduction}

In this paper we ask how likely it is for the universe to have
contained \linebreak primordial black holes.
We investigate universes which undergo a period of inflation
in their earliest stage, driven by a scalar field $\phi$ with a potential
$V(\phi)$ with a minimum $V(0) = 0$.
The results do not depend qualitatively on the exact form of the potential,
so for simplicity we consider a massive minimally coupled scalar
$V(\phi) = \frac{1}{2} m^2 \phi^2$.
The scalar field
starts out at a large initial value $\phi_0$ and acts as a cosmological
constant for some time until it reaches the minimum of its potential and
inflation ends.
We consider two different types of spacetimes: in the first, the
spacelike sections are simply 3-spheres and no black holes are present;
in the second, they have the topology $S^1\times S^2$, which is the
topology of the spatial section of the Schwarzschild-de~Sitter solution.
Thus these spaces can be interpreted as inflationary universes with a
pair of black holes.
In the inflationary period, the first type will be similar to a de~Sitter
universe, the second to a Nariai universe \cite{Nariai}.
To find the likelihood for primordial black holes, we assign probabilities
to both types of spacetimes using the Hartle-Hawking no-boundary proposal
(NBP)~\cite{HarHaw}.
This is the only proposal for the boundary conditions of the universe
that seems to give a well-defined answer in this situation.
It is not clear how to apply the so-called `tunneling proposal' in the
$S^1 \times S^2$ case.
If one takes the action to appear with the opposite sign as is done in
the $S^3$ case, one would reach the conclusion that a universe with a pair
of black holes was more likely than a universe without, and that the
probability would increase with the size of the black holes.
This is clearly absurd.

The NBP framework is summarized in Section \ref{secnbp}.
In Sections \ref{secdeSitter} and \ref{secNariai}
we review its implementation for cases
with a fixed cosmological constant.
In Section \ref{secs4} we introduce a massive scalar field and discuss the
solutions of the Euclidean Einstein equations for the $S^3$ case.
They will be slightly complex due to the time dependence of the effective
cosmological constant $(m\phi)^{-2}$.
We obtain the Euclidean action for those solutions.
In Section \ref{secs2} we go through a similar procedure
for the $S^1\times S^2$ case.
We find that the black hole grows during the inflationary period, a
noteworthy difference to the Nariai case with a fixed cosmological constant.
In Section \ref{secprob} we use the action to estimate the relative
probability of the two types of universes.
We find that black holes are suppressed for all but very large initial
values of $\phi_0$.

\sect{The Wave Function of the Universe} \label{secnbp}

The Hartle-Hawking no-boundary proposal states that the wave function
of the universe is given by
\beq
\Psi_0[h_{ij}, \Phi_{\partial M}] =
\int D(g_{\mu\nu}, \Phi) \, \exp \left[ -I(g_{\mu\nu}, \Phi) \right],
\eeq
where $(h_{ij}, \Phi_{\partial M})$ are the 3-metric and matter field
on a spacelike boundary $\partial M$ and the path integral is taken over
all compact Euclidean four geometries $g_{\mu\nu}$
that have $\partial M$ as their
only boundary and matter field configurations $\Phi$ that are regular on them;
$I(g_{\mu\nu}, \Phi)$ is their action.

The gravitational part of the action is given by
\beq
I_E = -\frac{1}{16\pi} \int_M d^4\!x\, g^{1/2}(R-2\Lambda)
      -\frac{1}{8\pi}  \int_{\partial M} d^3\!x\, h^{1/2} K,
\eeq
where $R$ is the Ricci-scalar, $\Lambda$ is the cosmological constant,
and $K$ is the trace of $K_{ij}$, the second fundamental form of the
boundary $\partial M$ in the metric $g$. For the origin of the boundary
term, see, e.\ g., ref.\ \cite{HawEQG}.

In the standard 3+1 decomposition \cite{ADM}, the metric is written as
\beq
ds^2 = N^2 d\tau^2 + h_{ij} (dx^i + N^i d\tau) (dx^j + N^j d\tau).
\eeq
Assuming that the NBP is satisfied at $\tau=0$, the Euclidean action then
takes the form
\[
I_E = -\frac{1}{16\pi} \int_{\tau=0}^{\tau_{\partial M}} N d\tau
      \int d^3\!x\, h^{1/2} (-K_{ij} K^{ij} + K^2 +\, ^3\!R - 2\Lambda)
\]
\beq
      +\frac{1}{8\pi}  \int_{\tau=0} d^3\!x\, h^{1/2} K.
\label{eqaction3+1}
\eeq
Here $^3\!R$ is the scalar curvature of the surface, and tensor operations
are carried out with respect to the surface metric $h_{ij}$.
In the first term the boundary terms are implicitly
subtracted out at $\tau=0$ and
$\tau=\tau_{\partial M}$. But it is an essential prescription of the NBP
that there {\em is no} boundary at $\tau=0$.
So the second term explicitly adds the contribution from $\tau=0$ back in.
It vanishes for universes with spacelike sections of topology $S^3$,
but can be non-zero for the topology $S^1 \times S^2$.

There are unresolved questions on how to choose the integration contour and
make the integral converge \cite{Luoko}, but we shall not discuss them here.
Instead, we will use the semiclassical approximation
\beq
\Psi_0[h_{ij}, \Phi_{\partial M}] \approx \sum_{n} A_n e^{-I_n},
\eeq
where the sum is over the saddlepoints of the path integral, i.\ e.\ the
solutions of the Euclidean Einstein equations.
In this paper, we neglect the prefactors $A_n$ and take only one saddlepoint
into account for a given argument of the wave function.
So the probability measure will be
\beq
\left| \Psi_0[h_{ij}, \Phi_{\partial M}] \right| ^ 2 =
\left| e^{-I} \right| ^ 2 =
e^{-2I^{\rm Re}},
\eeq
where $I^{\rm Re}$ is the real part of the Euclidean saddlepoint action.

By considering only spaces of high symmetry (homogeneous $S^3$ or
$S^1\times S^2$ spacelike sections) we restrict the degrees of freedom
in the metric to a finite number $q^{\alpha}$.
The Euclidean action for such a minisuperspace model with bosonic matter
will typically have the form
\beq
I = - \int N d\tau \left[ \frac{1}{2} f_{\alpha \beta}
    \frac{dq^{\alpha}}{d\tau} \frac{dq^{\beta}}{d\tau} +
    U \left( q^{\alpha} \right) \right].
\eeq
The saddlepoints will in general be complex solutions $q^{\alpha}(\tau)$
in the $\tau$-plane.
In the semiclassical approximation the following
relations for the real and imaginary parts of the saddlepoint actions hold:
\beq
-\frac{1}{2} \left( \nabla I^{\rm Re} \right) ^2 + \frac{1}{2}
\left( \nabla I^{\rm Im} \right) ^2 + U(q^{\alpha}) = 0 \label{eqchj}
\eeq
\beq
\nabla I^{\rm Re} \cdot \nabla I^{\rm Im} = 0,
\eeq
where the gradient and the dot product are both with respect to
$f^{\alpha\beta}$.
Therefore $I^{\rm Im}$ will be a solution of the
Lorentzian Hamilton-Jacobi equation
in regions of minisuperspace where $\Psi$ has the property that
\beq
\left( \nabla I^{\rm Re} \right) ^2 \ll \left( \nabla I^{\rm Im} \right) ^2
        \label{eqreim} ,
\eeq
This allows us to reintroduce a concept of Lorentzian time in such regions:
We find the integral curves of $\nabla I^{\rm Im}$ in minisuperspace and
define the Lorentzian time $t$ as the parameter naturally associated
with them.
Reversely, if we demand that the NBP should predict classical Lorentzian
universes at sufficiently late Lorentzian time, condition~\rf{eqreim}
must be satisfied.
This means that there must be saddlepoint solutions for which the path in the
$\tau$-plane can be deformed such that it is eventually almost parallel to the
imaginary $\tau$ axis and that all the $q^{\alpha}$ should be virtually
real at late Lorentzian times. In summary,
the following conditions must be met:
\begin{enumerate}
\item The NBP must be satisfied at $\tau$ = 0.
\item At the endpoint $\tau_{\partial M}$ of the path, the $q^{\alpha}$
      must take on the real values $q^{\alpha}_{\partial M}$ of the
      arguments of the wavefunction:
      \beq
      q^{\alpha}(\tau_{\partial M}) = q^{\alpha}_{\partial M}.
      \eeq
\item The $q^{\alpha}$ must remain nearly real in the Lorentzian vicinity
      of the endpoint:
      \beq
      \re \left( \left. \frac{dq^{\alpha}}{d\tau} \right|_{\tau_{\partial M}}
      \right) \approx 0.
      \eeq
\end{enumerate}


\sect{The de Sitter Spacetime} \label{secdeSitter}

In this and the next section we review vacuum solutions of the
Euclidean Einstein equations with a cosmological constant $\Lambda$.
First we look for a solution with spacelike sections $S^3$.
Therefore we choose the metric ansatz
\beq
ds^2 = N(\tau)^2 d\tau^2 + a(\tau)^2 d\Omega_3^2.
\eeq
The Euclidean action is
\beq
I = -\frac{3\pi}{4} \int N d\tau a \left( \frac{\da^2}{N^2} + 1 -
      \frac{\Lambda}{3} a^2 \right),
\eeq
A dot ($\,\dot{ }\,$) denotes differentiation with respect to $\tau$.
We define
\beq
H = \sqrt{\frac{\Lambda}{3}}.
\label{eqHLS4}
\eeq
Variation of $a$ and $N$ yields the equation of motion
\beq
\frac{\dda}{a} + H^2 = 0
\label{eqmLS4}
\eeq
and the Hamiltonian constraint
\beq
\frac{\da^2}{a^2} - \frac{1}{a^2} + H^2 = 0
\label{eqconstrLS4}
\eeq
in the gauge $N=1$.
A solution of equations~\rf{eqmLS4} and~\rf{eqconstrLS4} is given by
\beq
a(\tau) = H^{-1} \sin H\tau.
\label{eqsolLS4}
\eeq
It is called the de Sitter spacetime.
The NBP is satisfied at $\tau=0$, where $a=0$ and $\frac{da}{d\tau}=1$.
If we choose a path along the $\tau^{\rm Re}$-axis to $\tau=\frac{\pi}{2H}$,
the solution will describe half of the Euclidean de Sitter instanton $S^4$.
Choosing the path to continue parallel to the $\tau^{\rm Im}$-axis, $a(\tau)$
remains real and the conditions {\em i} to {\em iii} of the previous
section will be fulfilled:
\beq
\left. a(\tau^{\rm Im}) \right|_{\tau^{\rm Re}=\frac{\pi}{2H}} =
H^{-1} \cosh H\tau^{\rm Im}.
\eeq
This describes half of an ordinary Lorentzian de Sitter universe.

So with the above choice of path, equation~\rf{eqsolLS4} corresponds to
half of a real Euclidean 4-sphere joined to a real Lorentzian hyperboloid
of topology $R^1 \times S^3$. 
It can be matched to any $a_{\partial M} > 0$ by choosing the endpoint
appropriately, and for $a_{\partial M} > H^{-1}$ the wavefunction
oscillates and a classical Lorentzian universe is predicted.

The real part of the action for this saddlepoint is
\beq
I_{\rm de Sitter}^{\rm Re} =
  \frac{3\pi}{2} \int_0^{\frac{\pi}{2H}} d\tau^{\rm Re} a \left(
  H^2 a^2 - 1 \right)
  = -\frac{3\pi}{2\Lambda}.
\eeq
The Lorentzian segment of the path only contributes to $I^{\rm Im}$.

\sect{The Nariai Spacetime} \label{secNariai}

We still consider vacuum solutions of the
Euclidean Einstein equations with a cosmological constant, but we now
look for solutions with spacelike sections $S^1 \times S^2$.
The corresponding ansatz is the Kantowski-Sachs metric
\beq
ds^2 = N(\tau)^2 d\tau^2 + a(\tau)^2 dx^2 + b(\tau)^2 d\Omega_2^2.
\eeq
The Euclidean action is
\beq
I = -\pi \int N d\tau\, a \left( \frac{\db^2}{N^2}
+ 2 \frac{b}{a} \frac{\da \db}{N^2} + 1 -
\Lambda b^2 \right) + \pi \left[ -\da b^2 - 2 a b \db \right]_{\tau=0},
\label{eqactnariai}
\eeq
where the second term is the surface term of equation \rf{eqaction3+1}.
We define
\beq
H = \sqrt{\Lambda}.
\label{eqHLS2}
\eeq
Variation of $a$, $b$ and $N$ gives the equations of motion and the
Hamiltonian constraint:
\beqa
\frac{\ddb}{b} - \frac{\da\db}{ab}                           & = & 0
\\
\frac{\ddb}{b} + \frac{\da\db}{ab} + \frac{\dda}{a} + H^2    & = & 0
\\
2\frac{\da\db}{ab} + \frac{\db^2}{b^2} - \frac{1}{b^2} + H^2 & = & 0.
\eeqa
A solution is given by
\beq
a(\tau) = H^{-1} \sin H\tau,\: b(\tau) = H^{-1} = {\rm const}.
\eeq
It is called the Nariai spacetime.
The NBP is satisfied at $\tau=0$, where
\beq
a=0,\: \da=1,\: b=b_0\: {\rm and}\: \db=0.
\eeq
(There is a second way of satisfying the NBP for the Kantowski-Sachs metric
\cite{LaflPhD}, but it will not lead to a universe containing black holes.)
The path along the $\tau^{\rm Re}$-axis describes half of the Euclidean Nariai
instanton $S^2 \times S^2$. Both 2-spheres have the radius $H^{-1}$.
Continuing parallel to the $\tau^{\rm Im}$-axis, the solution remains real:
\beq
\left. a(\tau^{\rm Im}) \right|_{\tau^{\rm Re}=\frac{\pi}{2H}} =
H^{-1} \cosh H\tau^{\rm Im}\! ,\:
\left. b(\tau^{\rm Im}) \right|_{\tau^{\rm Re}=\frac{\pi}{2H}} =
H^{-1}.
\eeq
This describes half of a Lorentzian Nariai universe.
Its spacelike sections can be visualized as 3-spheres of radius $a$ with
a ``hole'' of radius $b$ punched through the North and South pole.
This gives them the topology of $S^1 \times S^2$.
Their physical interpretation is that of 3-spheres containing two black holes
at opposite ends.
The black holes have the radius $b$ and accelerate away from each other
as $a$ grows.
The Nariai universe is a degenerate case of the Schwarzschild-de~Sitter
spacetime, with the black hole horizon and the cosmological horizon having
equal radius~\cite{GinPer83}.

The above path corresponds to
half of a 2-sphere joined to a two- \linebreak dimensional hyperboloid at its
minimum radius $H^{-1}$, cross a 2-sphere of constant radius $H^{-1}$.
It can be matched to any $a_{\partial M} > 0$ but only to $b_{\partial M}
= H^{-1}$ so the wavefunction will be highly peaked around that value of $b$.

The first term of equation \rf{eqactnariai} vanishes and so
the real part of the action for the Nariai solution comes entirely
from the second term:
\beq
I_{\rm Nariai}^{\rm Re} = -\pi b_0^2 = -\frac{\pi}{\Lambda}.
\eeq

Now we compare the probability measures corresponding to the de Sitter and
Nariai solutions.
We find that in these models with a fixed cosmological constant
primordial black holes are strongly suppressed, unless $\Lambda$ is at least
of order $1$ in Planck units:
\beq
\exp\left( -2 I^{\rm Re}_{\rm Nariai} \right) =
\exp\left( \frac{2\pi}{\Lambda} \right) \ll
\exp\left( \frac{3\pi}{\Lambda} \right) =
\exp\left( -2 I^{\rm Re}_{\rm de Sitter} \right).
\eeq

\sect{An Inflationary Model Without Black Holes} \label{secs4}

Of course, we know that $\Lambda \approx 0$, and therefore the
models of the previous section are rather unrealistic.
However, in inflationary cosmology it is assumed that the very early
universe underwent a period of exponential expansion.
It has proven very successful to model this behaviour by introducing
a massive scalar field $\Phi$ with a potential $\frac{1}{2} m^2 \Phi^2$.
If this field is sufficiently far from equilibrium at the beginning of the
universe, the corresponding energy density acts like a cosmological constant
until the field has reached its minimum and starts oscillating.
During this time the universe behaves much like the Lorentzian de Sitter or
Nariai universes described above.

But there are two important differences
due to the time dependence of the effective cosmological constant
$\Lambda_{\rm eff}$:
Firstly, for the solutions of the Euclidean Einstein equations in the
complex $\tau$-plane one can no longer find a path on which the
minisuperspace variables are always real.
However, we shall see that it is possible to satisfy conditions
{\em i} to {\em iii} of Section \ref{secnbp} by choosing appropriate
complex initial values.
Secondly, it
will be found in the next section that
the black hole radius $b$ is no longer constant during inflation.

In this section, we introduce the massive scalar field for the model
corresponding to de Sitter spacetime, where the spacelike slices are
3-spheres containing no black holes.
This model was first put forward by Hawking \cite{Haw}. From
the fluctuations in the cosmic microwave background as
measured by COBE \cite{COBE} it follows
that $m$ is small compared to the Planck mass \cite{HalHaw}:
\beq
m \approx O \left( 10^{-5} \right).
\eeq
We will find complex solutions and the complex initial value
of the scalar field, and we calculate the real part of the action.
This has been done before by Lyons \cite{Lyons}, but his paper contains
a logical error to which we will come back later.

The ansatz for the Euclidean metric is again
\beq
ds^2 = N(\tau)^2 d\tau^2 + a(\tau)^2 d\Omega_3^2.
\eeq
Using the rescaled field
\beq
\phi^2 = 4 \pi \Phi^2
\eeq
we obtain the Euclidean action
\beq
I = -\frac{3\pi}{4} \int N d\tau a \left( \frac{\da^2}{N^2} + 1 -
    \frac{1}{3} a^2 \frac{\dphi^2}{N^2} - \frac{1}{3} a^2 m^2 \phi^2
    \right),
\eeq
so that the effective cosmological constant is
\beq
\Lambda_{\rm eff}(\tau) = m^2 \phi(\tau)^2.
\eeq
In analogy to equation \rf{eqHLS4} we define
\beq
H(\tau) = \sqrt{\frac{\Lambda_{\rm eff}(\tau)}{3}}
        = \frac{m \phi(\tau)}{\sqrt{3}}.
\eeq
Variation with respect to $a$, $\phi$ and $N$ gives the Euclidean equations of
motion and the Hamiltonian constraint:
\beqa
\frac{\dda}{a} + \frac{2}{3} \dphi^2 + \frac{1}{3} m^2 \phi^2    & = & 0
\label{eqELS4a}
\\
\ddphi + 3 \frac{\da}{a} \dphi - m^2 \phi                        & = & 0
\label{eqELS4phi}
\\
\frac{\da^2}{a^2} -\frac{1}{a^2} - \frac{1}{3} \dphi^2 +
\frac{1}{3} m^2 \phi^2                                           & = & 0.
\label{eqconstrS4}
\eeqa
To evaluate $\Psi_0(a_{\partial M}, \phi_{\partial M})$ using a semiclassical
approximation we must find solutions in the complex $\tau$-plane that meet
conditions {\em i} to {\em iii} of Section \ref{secnbp}.
In particular, the NBP must be satisfied:
\beq
a = 0,\: \da = 1,\: \phi = \phi_0\: {\rm and}\: \dphi = 0\; {\rm for}\:
\tau=0.
\label{eqNBPS4}
\eeq

Assume that the initial value of the scalar field is large and nearly real:
\beq
\phi_0^\re \gg 1 \gg \phi_0^\im.
\eeq
An approximate solution near the origin is given by
\beqa
a_{\cal I}(\tau)    & = & \frac{1}{H_0^\re} \sin H_0^\re \tau
\label{eqIAS4a}
\\
\phi_{\cal I}(\tau) & = & \phi_0 + \sum_{n=1}^{\infty} \frac{1}{n!}
                          \gamma_n \tau^n
\label{eqIAS4phi}
\\
&&{\rm for}\: \left| \tau \right| < O \left( 1/H_0^\re \right),
\nonumber
\eeqa
where the Taylor series is obtained by solving equation \rf{eqELS4phi}
iteratively for $\ddphi$, using the NBP conditions \rf{eqNBPS4} and
the approximation \rf{eqIAS4a} for $a$.
It has the property that
\beq
\gamma_{2n+1} = 0\;\; {\rm for\ all\ } n.
\label{eqgammaS4}
\eeq
We call equations \rf{eqIAS4a} and \rf{eqIAS4phi} the {\em inner
approximation}.
Writing down the Taylor expansion explicitly to lowest non-trivial order
\beq
\phi(\tau) = \phi_0 \left[ 1 + \frac{3}{8 \phi_0^2} \left( H_0 \tau
             \right) ^2 \right] + O(\tau^4)
\label{eqTaylorS4}
\eeq
shows that $\phi$ is almost constant near the origin.

As an {\em outer approximation} we use:
\beqa
\phi_{\cal O}(\tau) & = & \psi_0 + \frac{im}{\sqrt{3}} \tau +
                         \chi_0 \exp (3iH_0 \tau)
\label{eqOAS4phi}
\\
a_{\cal O}(\tau)    & = & a_0 \exp \left[ -\frac{im}{\sqrt{3}} \int_0^{\tau}
                 \phi(\tau') d\tau' \right] +
                 c_0 \exp \left[ \frac{im}{\sqrt{3}} \int_0^{\tau}
                 \phi(\tau') d\tau' \right]
\label{eqOAS4a}
\\
           &   & {\rm for}\; 0 < \tau^{\im} \ll \frac{\sqrt{3} \phi_0^\re}{m}.
\nonumber
\eeqa
While this solution does not satisfy the NBP, it will be good
outside the validity of the inner approximation.
Both the $\chi_0$-term and the $c_0$-term can be neglected for $\tau^{\im}
\gg 1/H_0^\re$, but they are useful for matching $a_{\cal O}$ and
$\phi_{\cal O}$ to $a_{\cal I}$
and $\phi_{\cal I}$ at some $|\tau| \approx O(1/H_0^\re)$.
Comparison with equation \rf{eqTaylorS4} shows that
\beq
\chi_0 \approx O \left( \frac{\sqrt{3}}{\phi_0^{\re}} \right) ,\:
\im (\chi_0) \approx 0.
\label{eqchiS4}
\eeq

In the region of the inner approximation,
$a$ will be nearly real on the Lorentzian line
$\tau^\re = \frac{\pi}{2H_0^\re}$.
Matching $a_{\cal O}$ to $a_{\cal I}$ fixes
\beq
a_0 \approx \frac{i}{2H_0^\re},\;
c_0 \approx \frac{-i}{2H_0^\re}
\eeq
and ensures that $a$ will remain nearly real on this line.
To make $\phi (\tau)$ roughly real on the same line,
by equations \rf{eqOAS4phi} and \rf{eqchiS4} we have to choose
\beq
\psi_0^\im = -\frac{\pi}{2\phi_0^\re}
\label{eqpsi0ImS4}
\eeq
in the outer approximation.

$\phi_0^\im$ in turn is fixed by matching $\phi_{\cal I}$ to $\phi_{\cal O}$.
Since it is very small, this requires evaluation of equation \rf{eqIAS4phi}
to a very high order $n$.
However, we need not calculate any coefficients since, by equation
\rf{eqgammaS4}, $\phi_{\cal I}^\im$ is constant along the imaginary axis
to any order $n$:
\beq
\phi_{\cal I}^\im\!\! \left. ( \tau^\im ) \right|_{\tau^\re = 0} = \phi_0^\im.
\label{eqIAconstS4}
\eeq
Therefore it is convenient to choose a matching point $\tau_M$
on the imaginary axis:
\beq
\tau_M^\re = 0,\; \tau_M^\im = O \left( 1/H_0^\re \right).
\eeq
By equations \rf{eqOAS4phi} and \rf{eqchiS4} $\phi_{\cal O}^\im$ is also
constant along this axis:
\beq
\phi_{\cal O}^\im\!\! \left. ( \tau^\im ) \right|_{\tau^\re = 0} = \psi_0^\im,
\label{eqOAconstS4}
\eeq
so the result of the matching analysis will be independent of the precise
choice of $\tau_M$ on the axis, as it should be.
The matching condition is
\beq
\phi_{\cal I}^\im(\tau_M) = \phi_{\cal O}^\im(\tau_M)
\eeq
and by equations \rf{eqpsi0ImS4}, \rf{eqIAconstS4} and \rf{eqOAconstS4}
we obtain
\beq
\phi_0^\im = \psi_0^\im = -\frac{\pi}{2\phi_0^\re}.
\eeq
This result is
non-trivial (e.\ g.\ $\phi_0^\re \neq \psi_0^\re$).
We now see why the correct value for $\phi_0^\im$ is obtained in ref.\
\cite{Lyons}, although actually only $\psi_0^\im$ is calculated there.

We have thus satisfied condition {\em ii} of Section \ref{secnbp}.
By the continuity of the outer approximation, condition {\em iii}
can be satisfied by fine-tuning $\phi_0^\im$.
Condition {\em i} is satisfied by the construction of the inner
approximation.
The only freedom left is the choice of $\phi_0^\re$.
This variable parametrizes the set of solutions.

To calculate the Euclidean action for the solutions given above,
we consider a path going along the real $\tau$-axis from the origin to
$\tau^\re = \frac{\pi}{2H_0^\re}$ and then
parallel to the imaginary $\tau$-axis
to $\tau_{\partial M}$. Both $a$ and $\phi$ are nearly real on the Lorentzian
segment of this path, so the real part of the action can be approximated
by an integral only over the first segment, using the inner approximation
\cite{Lyons}:
\beq
I_{S^3}^\re \approx \frac{3\pi}{2} \int_0^{\pi/2H_0^\re} d\tau^\re\, a_{\cal I}
              \left( \frac{1}{3} a_{\cal I}^2 m^2 \phi_{\cal I}^2 -1 \right)
      \approx -\frac{3\pi}{2 m^2 ( \phi_0^\re )^2}.
\eeq

The outer approximation is not valid after inflation ends, when
$\phi \approx 0$. However, at this point we are already well inside the
classical regime.
A dust phase will ensue where $\phi$ oscillates; $a$ and $\phi$ will both
remain real.
Approximate solutions for this regime have been given by Hawking and Page~
\cite{HawPag88}.

\sect{An Inflationary Model With Black Holes} \label{secs2}

We now introduce a massive scalar field on a universe with spacelike
sections $S^1 \times S^2$.
Thus we will obtain a cosmological model similar to the Nariai universe of
Section \ref{secNariai}.
We find the complex solutions, initial conditions and the action in analogy
to the previous section, but point out a few differences.

Again we use the Kantowski-Sachs metric
\beq
ds^2 = N(\tau)^2 d\tau^2 + a(\tau)^2 dx^2 + b(\tau)^2 d\Omega_2^2
\eeq
and the rescaled field
\beq
\phi^2 = 4 \pi \Phi^2.
\eeq
The Euclidean action is
\[
I = -\pi \int N d\tau\, a \left( \frac{\db^2}{N^2}
    + 2 \frac{b}{a} \frac{\da \db}{N^2} + 1 -
    b^2 \frac{\dphi^2}{N^2} - b^2 m^2 \phi^2 \right)
\]
\beq
    + \pi \left[ -\da b^2 - 2 a b \db \right]_{\tau=0},
\label{eq}
\eeq
and like in the previous section the effective cosmological constant is
given by
\beq
\Lambda_{\rm eff}(\tau) = m^2 \phi(\tau)^2.
\eeq
In analogy to equation \rf{eqHLS2} we define
\beq
H(\tau) = \sqrt{\Lambda_{\rm eff}(\tau)}
        = m \phi(\tau).
\label{eqHdefS2}
\eeq
Variation with respect to $a$, $b$, $\phi$ and $N$ gives the Euclidean
equations of motion and the Hamiltonian constraint:
\beqa
\frac{\ddb}{b} - \frac{\da\db}{ab} + \dphi^2                 & = & 0
\\
\frac{\ddb}{b} + \frac{\da\db}{ab} + \frac{\dda}{a} +
\dphi^2 + m^2 \phi^2                                         & = & 0
\\
\ddphi + \left( \frac{\da}{a} + 2 \frac{\db}{b} \right)
\dphi - m^2 \phi                                             & = & 0
\\
2\frac{\da\db}{ab} + \frac{\db^2}{b^2} - \frac{1}{b^2}
- \dphi^2 + m^2 \phi^2                                       & = & 0.
\eeqa
The NBP conditions corresponding to an instanton of topology
$S^2 \times S^2$ are:
\beq
a = 0,\: \da = 1,\: b=b_0,\: \db=0,\:
\phi = \phi_0\: {\rm and}\: \dphi = 0\; {\rm for}\:
\tau=0.
\label{eqNBPS2}
\eeq

With the new definition \rf{eqHdefS2} of $H$ the {\em inner approximation}
is given by:
\beqa
a_{\cal I}(\tau)    & = & \frac{1}{H_0^\re} \sin H_0^\re \tau
\label{eqIAS2a}
\\
\phi_{\cal I}(\tau) & = & \phi_0 + \sum_{n=1}^{\infty} \frac{1}{n!}
                          \gamma_n \tau^n
\label{eqIAS2phi}
\\
b_{\cal I}(\tau)   & = & \frac{1}{m \phi_{\cal I}(\tau)}
\label{eqIAS2b}
\\
&&{\rm for}\: \left| \tau \right| < O \left( 1/H_0^\re \right).
\nonumber
\eeqa
The {\em outer approximation} is:
\beqa
\phi_{\cal O}(\tau) & = & \psi_0 + i m \tau + \chi_0 \exp (iH_0 \tau)
\label{eqOAS2phi}
\\
a_{\cal O}(\tau)    & = & a_0 \exp \left[ -i m \int_0^{\tau}
                 \phi(\tau') d\tau' \right] +
                 c_0 \exp \left[ i m \int_0^{\tau}
                 \phi(\tau') d\tau' \right]
\label{eqOAS2a}
\\
b_{\cal O}(\tau)   & = & \frac{1}{m \phi_{\cal O}(\tau)}
\label{eqOAS2b}
\\
           &   & {\rm for}\; 0 < \tau^{\im} \ll \frac{\phi_0^\re}{m}.
\nonumber
\eeqa

A matching analysis completely analogous to that of the previous section shows
that $a$, $b$ and $\phi$ will be nearly real on the Lorentzian line
$\tau^\re = \frac{\pi}{2H_0^\re}$,
if we choose the following initial values:
\beq
\phi_0^\im = -\frac{\pi}{2\phi_0^\re},\:
b_0 = \frac{1}{m \phi_0};
\eeq
$\phi_0^\re$ is a free parameter.

An interesting feature of the outer approximation is that the black hole
radius grows with the horizon scale during inflation.
On the Lorentzian line $\tau^\re = \frac{\pi}{2H_0^\re}$ the field decreases
linearly with time until it reaches zero
and inflation ends.
By equations \rf{eqOAS2phi} and
\rf{eqOAS2b} $b$ becomes very large on the timescale
\beq
\Delta \tau_{\rm growth} = \frac{\phi_0^\re}{m}.
\label{eqBHgrowth}
\eeq

Again the inner approximation is used to calculate
the real part of the Euclidean action.
As in Section \ref{secNariai} it comes entirely from the $\tau=0$ term:
\beq
I_{S^1 \times S^2}^{\rm Re} \approx -\pi \left( b_0^\re \right) ^2 \approx
-\frac{\pi}{m^2 \left( \phi_0^\re \right) ^2}.
\eeq

\sect{The Probability for Primordial Black Holes} \label{secprob}

In the previous two sections we have calculated the action for two
inflationary universes.
We now compare the corresponding probability measures
\beq
P_{S^3}(\phi_0^\re) = \exp \left( \frac{3\pi}{m^2 \left( \phi_0^\re \right)
^2} \right)\; {\rm and}\;
P_{S^1 \times S^2}(\phi_0^\re) =
\exp \left( \frac{2\pi}{m^2 \left( \phi_0^\re \right) ^2} \right).
\eeq
The universe containing black holes
is heavily suppressed,
if $\phi_0^{\rm Re}$
is not large enough to make the initial effective cosmological
constant equal to the Planck value.
Thus the formation of black holes with initial sizes significantly larger
than the Planck scale is very unlikely.
The semi-classical approximation should be good in these situations, so
one can have confidence in this conclusion.

The semi-classical approximation will break down for solutions with
initial cosmological constants of the Planck value in a region where the
curvature is on the Planck scale.
However, this region contributes an action less than one in Planck units
and one would not expect quantum effects to change this.
Thus it seems clear that the only primordial black holes with any
significant probability start with no more than the Planck size:
\begin{equation}
b_0^{\rm Re} < O(1).
\end{equation}
This corresponds to a large initial value of the scalar field
\begin{equation}
\phi_0^{\rm Re} > O \left( 10^5 \right).
\end{equation}

The Nariai solution is unstable to quantum fluctuations~\cite{GinPer83}.
At the beginning of inflation
it becomes a non-degenerate Schwarzschild-de~Sitter spacetime.
Once the black hole horizon is inside the cosmological horizon
the black hole will start to lose mass due to Hawking radiation.
If the black hole horizon is somewhat smaller than the cosmological
horizon, the black hole will evaporate and disappear.
However, there is a significant probability that the areas of the two
horizons will be nearly enough equal for them to increase together.
The consequences of this result for the global structure of the universe
will be presented in a forthcoming paper.

\end{document}